\documentclass[aps,prb,showkeys,preprint,showpacs,12pt]{revtex4-1}
\usepackage{graphics,graphicx}
\usepackage{color}
\usepackage{amsbsy,amssymb}
\usepackage{amsmath}
\usepackage{epsfig}
\usepackage{color}
\usepackage{textcomp}
\begin{document}
\title{Ferromagnetic behavior of ultrathin manganese nanosheets}
%%%%%%%%%%%%%%%%%%%%%%%%%%%%%%%%%%%%%%%%%%%%%%%%%%%%%%%%%%%%%%%%%%%%%%%%%%%%%%%%%%%%%%%%%%%%

%%%%%%%%%%%%%%%%%%%%%%%%%%%%%%%%%%%%%%%%%%%%%%%%%%%%%%%%%%%%%%%%%%%%%%%%%%%%%%%%%%%%%%%%%%%
\author{Sreemanta Mitra$^{1,2}$}
\email[]{sreemanta85@gmail.com}
\author{Amrita Mandal$^{1,2}$}
%\author{Dhriti ranjan Saha$^{1}$}
\author{Anindya Datta $^{3}$}
%\email[]{anindya$_$datta@yahoo.com}
\author{Sourish Banerjee$^{2}$}
\author{Dipankar Chakravorty$^{1,\dag}$}
\email[]{mlsdc@iacs.res.in}
%%%%%%%%%%%%%%%%%%%%%%%%%%%%%%%%%%%%%%%%%%%%%%%%%%%%%%%%%%
%\author{Sreemanta Mitra$^{1,2}$,Amrita Mandal$^{1,2}$,Anindya Datta$^{1,3}$,Sourish Banerjee$^{2}$ and Dipankar Chakravorty$^{1,\dag}$}
\affiliation{
$^{1}$
 MLS Prof.of Physics' Unit,Indian Association for the Cultivation of Science, Kolkata-700032, India.\\ }
\affiliation{
$^{2}$
Department of Physics, University of Calcutta, Kolkata-700009, India.\\}
%\affiliation{
%$^{3}$ Department of Physics, M.U.C. Woman's College, Burdwan, India.\\}

\affiliation{
$^{3}$
University School of Basic and Applied Science (USBAS),Guru Govind Singh Indraprastha University,New Delhi, India\\}
%%%%%%%%%%%%%%%%%%%%%%%%%%%%%%%%%%%%%%%%%%%%%%%%%%%%%%%%%%%%%%%%%%%%%%%%%%%%%%%%%%%%%%%%%%%%%%
%\begin{document}
%%%%%%%%%%%%%%%%%%%%%%%%%%%%%%%%%%% Abstract %%%%%%%%%%%%%%%%%%%%%%%%%%%%%%%%%%%%%%%%%%%%%%%%%%%%%%%
\begin{abstract}
    Ferromagnetic behaviour has been observed experimentally for the first time in nanostructured Manganese.  
Ultrathin ($\sim$ 0.6 nm) Manganese nanosheets  have been synthesized inside the two dimensional channels of sol-gel derived Na-4 mica. 
The magnetic properties of the confined system are measured within 2K-300K temperature range. 
The confined structure is found to show a ferromagnetic behaviour with a nonzero coercivity value. 
The coercivity value remains positive throughout the entire temperature range of measurement.
The experimental variation of susceptibility as a function of temperature has been satisfactorily explained on the basis of a 
two dimensional system with a Heisenberg Hamiltonian involving direct exchange interaction.
\end{abstract}

\maketitle

\section{Introduction}\label{sec:1}
Manganese (Mn) is probably one of the most important dopants for semiconductors for creating ferromagnetic response ~\cite{ohnoprl,deitlsc,sharmanat,dongcm,xujpcc,norrisnl,
liujpcc}. 
In addition, dilute solution of Mn based systems has wide variety of magnetic properties ~\cite{chuprl,mcdonald}. However, Mn by itself is not ferromagnetic even though Mn
 is a\textquoteleft d\textquoteright shell based transition metal in its bulk form which provokes fundamental question about its ground state which is still not very clearly understood.
\par 
Unique properties of manganese among the first row transition metal elements as atoms, cluster or crystal arise probably due to its atomic configuration.
 Manganese has exactly half filled 3d orbital and fully filled 4s orbital $3d^{5}4s^{2}$ as electronic configuration. The energy required to change the electronic 
configuration from $3d^{5}4s^{2}$ to $3d^{6}4s^{1}$  is high enough ({$\sim$} 2.4 eV),~\cite{nayak} to keep this as its ground state configuration. 
 If two Mn atoms are brought closer,the 3d and 4s orbitals would split into bonding and antibonding states respectively. So the magnetic ground state (ferro or antiferro) 
would depend on energy of splitting and exchange interaction ~\cite{nayak}. The calculations showed that ferromagnetic ground state was not
 energetically stable, for Mn ~\cite{hobbs}. There is a great deal of debate among different theoretical studies, even for the simplest dimer molecule $Mn_{2}$,
 with regard to ferromagnetic as well as antiferromagnetic ground states ~\cite{nesbet,pederson,parvanova,harris,morisato,charles,wang,yamamoto,satadal}.
Spin density functional calculations of manganese nanostructures like nanowire or a nanorod,showed these to be in a high $-$ moment state with magnetic moments per atom 
having values in the range 2.96 $\mu_{B}$ to 3.79 $\mu_{B}$ depending on the morphology of the nanostructure \cite{martinprb}.   
 Experimentalists showed that small manganese clusters exhibited complex magnetic behaviour with signature of superparamagnetism ~\cite{markprl,markprb,mertes,baumann}. 
On the basis of theoretical understanding and experimental findings, magnetic property fluctuates between ferro and antiferro ground state as the 
cluster size changes from 40 to 80 atoms.Beyond a cluster size of 80 atoms the structures slowly converge to the bulk and magnetic ordering becomes antiferromagnetic. 
However, some magnetic deflection experiments showed non-zero magnetic moments for manganese clusters of 11 to 99 atoms, indicating ferromagnetic ordering of atomic spins,
 which had a lower limit of number of atoms in a cluster different from that mentioned earlier ~\cite{markprl}. So, manganese showed contradictory magnetic 
ground state from the stand point of theory as well as experiments. Induced ferromagnetism was observed in manganese on clean 
ferromagnetic substrates ~\cite{dreysse,yamada,brien} whereas for non magnetic substrate, monolayer deposition of Mn resulted in an
anti ferromagnetic ordering \cite{bode}. 1000 atom Mn clusters of height 10 nm and diameter 15-25 nm on Si(111) and Si(112) exhibit temperature dependent
ferromagnetic like behavior below 10 K,due to some surface orientation effect, which may be correlated to the surface dangling bond densities or cluster shape ~\cite{prokes}.
 However, for Mn thin films the reported ground state was antiferromagnetic in nature
~\cite{new}. Although two-dimensional Mn crystalline systems would be more stable compared to small clusters, its ferromagnetic response remains illusive till date.
 Hence the search for magnetic behavior in non-ferromagnetic transition metals has often focused on the effect of reduced dimension ~\cite{markprl}.
We have synthesized manganese nanosheets inside the two-dimensional crystal channels of Na-4 mica, with a very simple synthesis method. 
In this paper, we report on their ferromagnetic behaviour.
%%%%%%%%%%%%%%%%%%%%%%%%%%%%%%%%%%%%%%%%%%%%%%%
%%%%%%%%%%%%%%%%%%%%%%%%%%%%%%%%%%%%%%%%%%%%%%%%%%%%%%%%%%%%%%%%%%%%%%%%%%
\par
\section{Synthesis and Characterization}
Na-4 mica ($Na_{4}Mg_{6}Al_{4}Si_{4}F_{4}O_{20},xH_{2}O$)template
was synthesized by usual sol$-$gel technique, taking Aluminum Nitrate, Magnesium Nitrate,(used as obtained from E Merck) Tetraethylorthosilicate (TEOS),
and Ethanol as precursors .
%%%%%%%%%%%%%%%
In brief, in order to prepare 1 gm of Na-4 mica powder, firstly,2.186 gm Aluminum Nitrate,2.241 gm Magnesium Nitrate were dissolved in  ethanol, 
and then the solution was stirred vigorously for 1 hour to have a homogeneous mixture. To this 1.291 cc of TEOS was added to achieve the target composition.
0.1 N nitric acid was added as a catalyst. This solution was stirred for 3 hours and placed in an air oven at 333K for 3 days. The dried gel was 
crushed and calcined at 748K for 12 hours. Equal amount of gel powder and crystalline sodium fluoride (NaF) were mixed thoroughly and heated at 1163K for 18 hours in a platinum
crucible under ordinary atmosphere. This was needed for the dissolution of the gel in NaF. The reaction product was washed thoroughly with saturated boric acid
several times to remove the water insoluble fluoride salts. The powder was then washed again with 1M NaCl solution  3 times in order to 
completely saturate all exchange sites with $Na^{+}$ ions. The resultant product was then washed with deionized water several times, and dried at 333K in an air oven, 
to obtain pure Na-4 mica powder.
%%%%%%%%%%%%%%%%
 
The unit cell of Na-4 mica has layered structure with interlayer spacing of 0.6nm ~\cite{pkm,santanu,paulus}. 2.36 gm of  Na-4 mica powder was then subjected to ion 
exchange reaction $2Na^{+}\Leftrightarrow Mn^{2+}$ and soaking in a mixture of Manganese Nitrate[$Mn(NO_{3})_{2}$] and 
Dextrose ($C_{6}H_{12}O_{6}$) in aqueous solution at an elevated temperature (368 K) and pressure inside a teflon coated autoclave cell for five days.
 The pH of the solution was kept neutral throughout the ion exchange process.
The latter could occur only in the case of ions which were mobile. As such in the present case,ion exchange reactions with the other species viz; Al,Mg,Si are ruled out.
 The resultant powder was taken out of autoclave and washed thoroughly
with deionized water  several times to ensure that no Manganese Nitrate molecule was present on the surface of Na-4 mica powder. This was confirmed by a simple
chemical group test analysis. Sodium Carbonate was added to the filtrate and no white colored precipitate of Manganese Carbonate was obtained. The washed powder
was then put in an alumina boat and placed in a muffle furnace at 675 K under ordinary atmosphere for two hours. Carbon of Dextrose molecules reduced the Mn ion 
into Mn metal while becoming Carbon Dioxide. 
X-ray diffraction of the material was performed using a Bruker D-8 SWAX x-ray diffractometer with a Cu$K_{\alpha}$  monochromatic source of wavelength 0.15408 nm. 
In order to study the microstructure, Mn nanosheets were extracted from the mica channels, by etching the composite sample with $10{\%}$ HF aqueous solution and 
centrifuged in SORVALL RC 90 ultracentrifuge at 30,000 rpm for 30 min. The resultant residue was washed thoroughly with de-ionized water, and then dispersed in acetone. 
From that a drop was taken and mounted in a JEOL 2010 transmission electron microscope operated at 200 kV and investigated.Magnetic properties of the composite 
were studied by an MPMS SQUID magnetometer (Quantum Design) in the temperature range 2-300 K.
%%%%%%%%%%%%%%%%%%%%%%%%%%%%%%%%%%%%%%%%%%%%%%
\par
\section{Results \& Discussions}
Fig.1 shows the x-ray diffraction pattern of the composite sample in the range 2${\theta}$=5-80$^{o}$. The presence of manganese was confirmed by 
the standard JCPDS value (file number: 17-0910).Only (200) plane of manganese grows inside the two-dimensional mica nanochannels. 
Rest of the lines originate from the basic structure of Na-4 mica ~\cite{park,lee}. The c axis of Na-4 mica is found to be perpendicular to the $(200)$ plane 
of Mn allowing the growth along this direction only.
%%%%%%%%%%%%%%%%%%%%%%%%%%%%%%%%%%%%%%%%%%%%%%%%%%%%%%%%%%%
\begin{figure}
%\vskip 0.4 cm
%\centering
\includegraphics[width=8.25cm]{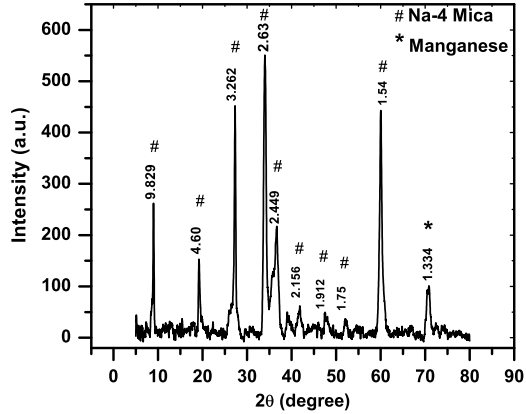}
\caption{ XRD pattern of Manganese and Na-4 mica composite.The numbers shown
indicate the corresponding interplanar spacing.}
\label{fig.1}
   \end{figure}    
%%%%%%%%%%%%%%%%%%%%%%%%%%%%%%%%%%%%%%%%%%%%%%%%%%%%%%%%%%%%%%%%%%%%%%%%%%%%%%
Fig.2 (a) shows the transmission electron micrograph of the randomly assembled partially etched nanocomposites containing Mn nanosheets.
 Fig.2(b) shows a zoomed view of that assembly. 
 The nanosheets are formed with manganese nanodiscs of circular and rhombus like structures. As the manganese was formed 
within the layers of Na-4 mica, its thickness is limited by the channel thickness,i.e. 0.6 nm. Fig. 2(c) shows the high resolution lattice image of one of the nanosheets 
where the $(200)$ plane of Mn was observed along with the lattice planes corresponding to Na-4 mica. The interplanar spacing was found to be 0.133 nm.
As a result the interatomic distance becomes 0.266 nm. From TEM pictures, the sizes of the nanosheets were found to be around 70-150 nm. Fig.2 (d) shows the
selected area electron diffraction (SAED) pattern of one of the nanosheets. As the sample was partially etched some spots in the SAED pattern corresponded to Na-4 mica 
structure, whereas, spot due to (200) plane of manganese arose as expected. The results are summarized in table.1. In order to delineate the thickness of the manganese nanosheet
synthesized in the present work, Atomic Force Microscope (Veeco model CP II) was used. The height profile of the nanosheets were obtained by etching the composite powder in 
$10{\%}$ HF aqueous solution for 4 days and dispersed on freshly cleaved atomically flat mica (supplied by SPI,USA) surface. A typical profile is shown in Fig.3. 
The heights measured are either 0.6nm or its integral multiple which confirms that the manganese nanosheets have indeed grown within the nano channels of Na-4 mica 
structure. The AFM images gave a thickness value equal to the thickness of interlayer space in Na-4 mica reported in the literature \cite{pkm,santanu,paulus}.
These facts lead us to conclude that the original films were not thicker than 0.6 nm and hence dissolution of Manganese during etching did not take place. 
%%%%%%%%%%%%%%%%%%%%%%%%%%%%%%%%%%%%%%%%%%%%%%%%%%%%%%%%%%%%%%%%%%%%
\begin{figure}
%\vskip 0.4 cm
%\centering
\includegraphics[width=8.25cm]{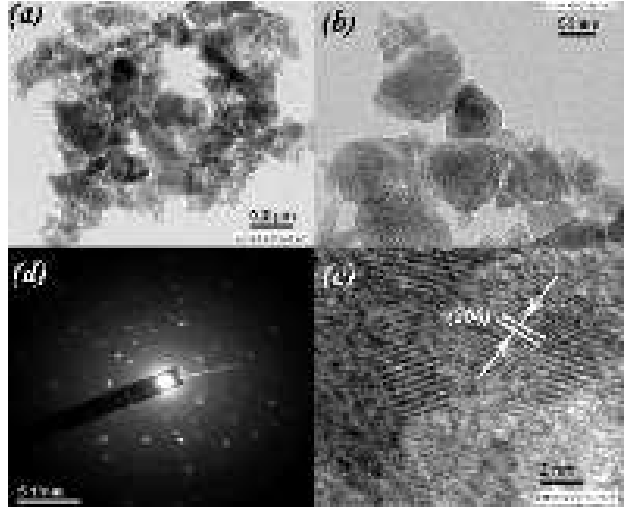}
\caption{(a)Transmission electron micrograph of manganese nanosheets, (b)A zoomed view of (a), (c) high resolution lattice image of one of manganese nanosheets, 
              (d) Selected Area electron Diffraction pattern of (b)}
\label{fig.2}
   \end{figure}
%%%%%%%%%%%%%%%%%%%%%%%%%%%%%%%%%%%%%%%%%%%%%%%%%%%%%%%%%%%%%%%%%%%%% 
%%%%%%%%%%%%%%%%%%%%%%%%%%%%%%%%%%%%%%%%%%%%%%%%%%%%%%%%%%%%%%%%%%%%% 
\begin{table}
\caption{Interplanar spacing estimated from electron diffraction data and JCPDS file respectively.(The numbers in the parenthesis describe the miller indices 
of the corresponding lattice planes.)}
\label{tab.1}
\begin{center}
\begin{tabular}{lcr}
 \hline
Observed & Na-4 mica & Manganese\\ (nm) & (nm) & (nm)\\
 \hline
0.30  & 0.303  (023)  &  $-$ \\
0.26  & 0.263  (20-1) & $-$ \\
0.20  & 0.202  (006)  &  $-$ \\
0.18  & 0.186  (20-5) & $-$ \\
0.17  & 0.173  (205)  &  $-$ \\
0.16  & 0.166 (-135)  & $-$ \\
0.141 & 0.143  (007)  & $-$ \\
0.133 & $-$             & 0.1336 (200) \\
\hline
\\
$-$ indicates that there exist
\\
 no planes with these interplanar spacings
\\
 in the phases concerned.
\end{tabular}
\end{center}
\end{table}
%%%%%%%%%%%%%%%%%%%%%%%%%%%%%%%%%%%%%%%%%%%%%%%%%%%%%%%%%%%%%%%%
%%%%%%%%%%%%%%%%%%%%%%%%%%%%%%%%%%%%%%%%%%%%%%
\begin{figure}
%\vskip 0.4 cm
%\centering
\includegraphics[width=8.25cm]{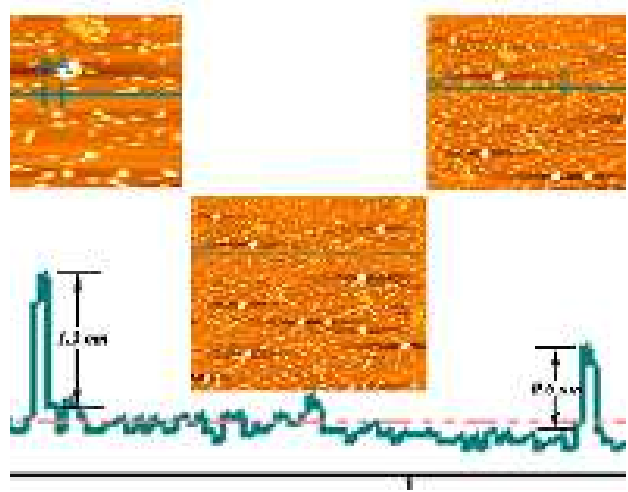}
\caption{(Color Online) AFM image of Manganese nanosheets and the height profile delineated thereof.}
\label{fig.3}
   \end{figure} 
%%%%%%%%%%%%%%%%%%%%%%%%%%%%%%%%%%%%%%%%%%%%%%%%%%%%%%%%%%%%%%
Magnetic measurements were carried out for the composite powder in the temperature range 2-300 K.  Fig.4 shows the variation of magnetization as a function of
temperature measured at an applied magnetic field 5 mT under both zero field cooled (ZFC) and field cooled (FC) conditions.
The absence of any local maxima in the ZFC magnetization - temperature curve indicates ferromagnetic coupling between the spins.
This is borne out by the magnetization - magnetic field hysteresis curve measured at 2 K, which is shown in Fig.5. In the inset of Fig.5,
the zoomed hysteresis curve near zero magnetization has been depicted and this shows a nonzero coercivity value ($\sim 60 $ Oe). 
%%%%%%%%%%%%%%%%%%%%%%%%%%%%%%%%%%%%%%%%%%%%%%%%%%%%%%%%%%%%%
\begin{figure}
%\vskip 0.4 cm
%\centering
\includegraphics[width=8.25cm]{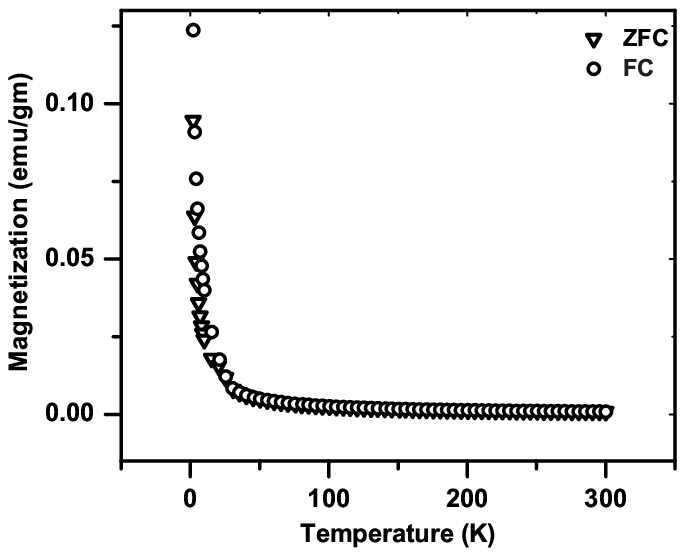}
\caption{Variation of magnetization ({\it M}) with temeperature ({\it T}) under both field cooled (FC) and Zero Field Cooled (ZFC) conditions measured at 5 mT.}
\label{fig.4}
   \end{figure}  
%%%%%%%%%%%%%%%%%%%%%%%%%%%%%%%%%%%%%%%%%%%%%%%%%%%%%%%%%%%%%%%%%%%%%%%%%%%
\begin{figure}
%\vskip 0.4 cm
%\centering
\includegraphics[width=8.25cm]{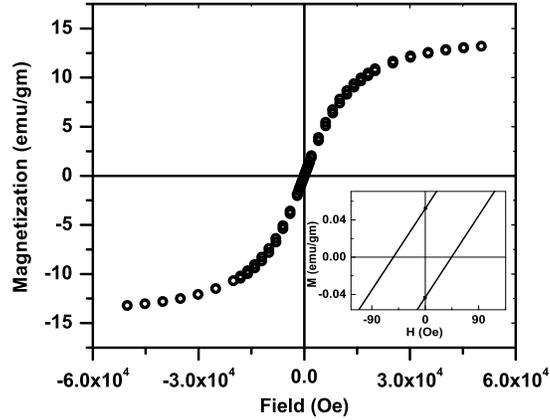}
\caption{ Variation of magnetization ({\it M}) with magnetic field ({\it H}) measured at 2K, (inset) zoomed view near zero magnetization.}
\label{fig.5}
\end{figure}
%%%%%%%%%%%%%%%%%%%%%%%%%%%%%%%%%%%%%%%%%%%%%%%%%%%%%%%%%%%%%%
The magnetization -magnetic field hysteresis curve for the composite at 300K is also shown in Fig.6.  A finite coercivity present
 here (as observed from the inset) also indicates a ferromagnetic behaviour even at the room temperature. The saturation magnetization was not achieved even at maximum field,
indicating all spins can not become parallel to the magnetic field due to the thermal energy. 
Fig.7 shows the magnetization as a function of magnetic field in the case of pure Na-4 mica,which shows a  
diamagnetic characteristic. We have presented our results after subtracting this contribution.
As regards the Curie temperature of the ultrathin Mn films we believe this to be above room temperature in view of the fact that at
room temperature a magnetic hysteresis was observed. The precise nature of size effect on the magnetic behaviour can not be inferred from our experimental data
because the latter pertain to only one thickness value of Mn nanosheet viz;0.6nm.
 Also, the EDAX data
(measured in Field emission scanning electron microscope (JEOL JSM-6700F).),in Fig.8 show that there is no magnetic 
impurity present in the  nanocomposite under investigation.
The EDAX data show that the amount of oxygen atoms present is just sufficient to fulfill the stoichiometric needs of Aluminium,Magnesium,
and Silicon present in the Na-4 mica phase. Hence, it can be safely concluded that there is no possibility of formation of oxides of manganese in our system.
This is also consistent with the fact that Mn was formed by subjecting it to reduction treatment at 675 K.
%%%%%%%%%%%%%%%%%%%%%%%%%%%%%%%%%%%%%%%%%%%%%%%%%%%
  \begin{figure}
%\vskip 0.4 cm
%\centering
\includegraphics[width=8.25cm]{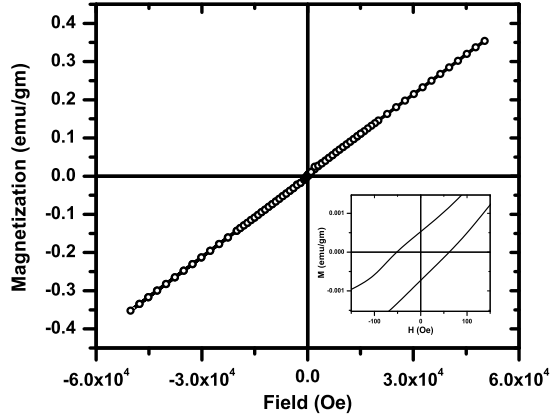}
\caption{ Variation of magnetization ({\it M}) with magnetic field ({\it H}) measured at 300K, (inset) zoomed view near zero magnetization.}
\label{fig.6}
\end{figure}
%%%%%%%%%%%%%%%%%%%%%%%%%%%%%%%%%%%%%%%%%%%%%%%%%%%%%%%%%%%%%%%%%%%%%%%%%%%
\begin{figure}
%\vskip 0.4 cm
%\centering
\includegraphics[width=8.25cm]{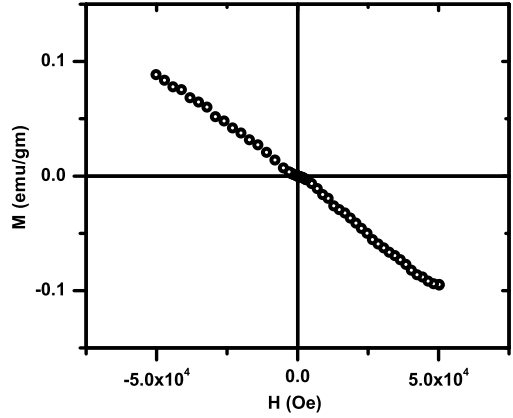}
\caption{ Variation of magnetization ({\it M}) with magnetic field ({\it H}) for Na-4 mica}
\label{fig.7}
\end{figure}
%%%%%%%%%%%%%%%%%%%%%%%%%%%
\begin{figure}
%\vskip 0.4 cm
%\centering
\includegraphics[width=8.25cm]{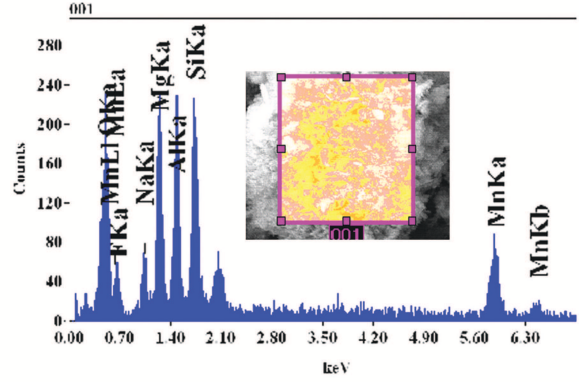}
\caption{(Color online) EDAX analysis of the nanocomposite}
\label{fig.8}
\end{figure}
%%%%%%%%%%%%%%%%%%%%%%%%%%%%%%%%%%%%%%%%%%%%%%%%%%%%%%%%%%%%%%%%%%%%%%%%%%%%%%%%
\par
In order to find out the possible origin of this ferromagnetism, out of two exchange
interactions possible viz; direct and indirect, we rule out the possibility of indirect exchange interaction, as the hopping integral decays
exponentially with the distance between magnetic centres, and stick to direct exchange interaction \cite{martinprb,satadal}. We have applied Heisenberg Hamiltonian
\begin{equation}
 H= - 2\sum_{<ij>} J_{ij}S_{i} . S_{j}
\end{equation}
\par
based variation of susceptibility with temperature for a two dimensional ferromagnet ~\cite{takahashi},
\begin{equation}
 \chi \sim exp (\beta /T)
\end{equation}
\par
with
\begin{equation}
 \beta=(4\pi JS^{2})/k_{B} 
\end{equation}
\par
where J is exchange coupling constant, S is spin quantum number, k$_B$ is Boltzman constant. The calculation has been done by taking $S=5/2$ for manganese and 
$(J/k_{B})$ as parameter. The fitting looks comprehensive and positive J value indicates ferromagnetic coupling. Both the experimental data 
and the fitted curve have been shown in Fig.9.   
%%%%%%%%%%%%%%%%%%%%%%%%%%%%%%%%%%%%%%%%%%%%%%%%%%%%%%%%%%%%%%%%%%%%%%%%%%%%%%%%%%%%%%

\begin{figure}
%\vskip 0.4 cm
%\centering
\includegraphics[width=8.25cm]{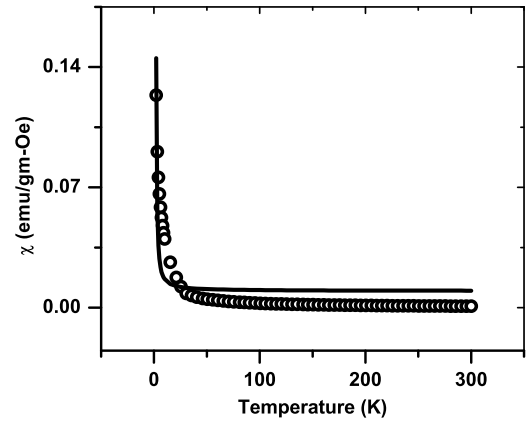}
\caption{Variation of susceptibility ($\chi$) with temeperature ({\it T}) (open circle: Experimental Data; solid line: Theoretically fitted curve).}
\label{fig.9}
   \end{figure}
%%%%%%%%%%%%%%%%%%%%%%%%%%%%%%%%%%%%%%%%%%%%%%%%%%%%%%%%%%%%%%%%%
A previous theoretical investigation did arrive at the conclusion that monolayers and bilayers of Mn grown on tungsten (100) plane 
have a ferromagnetic ground state \cite{ferriani}. Considering the thickness of 0.6 nm for our manganese films, our experimental results on their ferromagnetic
behaviour is consistent with the above mentioned theoretical calculations. Also theoretical prediction shows that the obtained interatomic distance of manganese atoms
ensures ferromagnetic response \cite{harris}.   
%%%%%%%%%%%%%%%%%%%%%%%%%%%%%%%%%%%%%%%%%%%%%%%%%%%%%%%%%%
 \par
\section{Conclusions}
In summary,we have synthesized manganese nanosheets of thickness 0.6 nm in the nanochannels of the layered Na-4 mica by a simple process. These nanosheets are made up of 
nanodiscs of size of about 70-150 nm. The nanodiscs consist of (200) planes of Mn,aligned parallel to the nanochannel itself,which was confirmed both 
by x-ray diffraction and selected area electron diffraction experiments. This confined Mn shows ferromagnetic behaviour in its ultrathin configuration 
as is evident from the magnetic measurements.Calculation of magnetic susceptibility variation as a function of temperature on the basis of Heisenberg 
ferromagnetic model involving direct exchange interaction, matches the experimental data satisfactorily.
%%%%%%%%%%%%%%%%%%%%%%%%%%%%%%%%%%%%%%%%%%%%%%%%%%%%%%%%%%%%%
%%%%%%%%%%%%%%%%%%%%%%%%%%%%%%%%%%%%%%%%
%\acknowledgement
\section*{Acknowledgement}
The work was supported by a grant awarded by Nano Mission Council, Department of Science and Technology, New Delhi. Sreemanta Mitra and Amrita Mandal thank 
University Grants Commission, New Delhi, for Jounior Research Fellowships. D.Chakravorty thanks Indian National Science Academy for awarding an 
Honorary Scientist's position. SM thanks Dr. Molly DeRaychaudhury for some fruitful discussions.Support was partly derived from a grant received 
from Department of Science and Technology, New Delhi, under an Indo-Australian Project on Nanocomposites. 
%%%%%%%%%%%%%%%%%%%%%%%%%%%%%%%%%%%%%%%%%%%%%%%%%%%%%%%%
%\bibliography{Bibliography}
%Merlin.mbs v4.21 2009-07-09.
%
\end{document}